\documentclass[12pt]{article}
\usepackage{amsthm,amsfonts}
\def\Box{\qed}
\title{Magnetic Lieb-Thirring inequalities with optimal dependence
on the field strength}
\author{L\'aszl\'o Erd\H os
 \thanks{Partially supported by NSF grant DMS-0200235
and by MSRI}
\\ School of Mathematics, GeorgiaTech and Maphysto \\
and \\
Jan Philip Solovej \thanks{Work partially supported
     by the Danish Natural Science Research Council, by MaPhySto -- Centre
for Mathematical
     Physics and Stochastics, funded by a
     grant from The Danish National Research Foundation, by the EU
research network HPRN-CT-2002-00277, and by MSRI.}
\\ Department of Mathematics, University of Copenhagen
\\
\\
{\it Dedicated to Elliott H. Lieb on his 70-th birthday}}

\date{\today}

\newtheorem{theorem}{Theorem}[section]
\newtheorem{proposition}[theorem]{Proposition}
\newtheorem{corollary}[theorem]{Corollary}
\newtheorem{lemma}[theorem]{Lemma}
\newtheorem{definition}[theorem]{Definition}

\newcommand{\rd}{{\rm d}}
\newcommand{\be}{\begin{equation}}
\newcommand{\ee}{\end{equation}}
\newcommand{\bey}{\begin{eqnarray}}
\newcommand{\eey}{\end{eqnarray}}
\newcommand{\beys}{\begin{eqnarray*}}
\newcommand{\eeys}{\end{eqnarray*}}

\newcommand{\bB}{{\bf B}}
\newcommand{\bA}{{\bf A}}
\newcommand{\bZ}{{\bf Z}}
\newcommand{\bp}{-i\nabla}

\newcommand{\bsigma}{\mbox{\boldmath $\sigma$}}
\newcommand{\bomega}{\mbox{\boldmath $\omega$}}

\newcommand{\bR}{{\bf R}}
\newcommand{\bC}{{\bf C}}

\newcommand{\bn}{{\bf n}}

\newcommand{\bv}{{\bf v}}

\newcommand{\e}{\varepsilon}
\newcommand{\tD}{\widetilde \D}

\renewcommand{\a}{\alpha}

\newcommand{\Tr}{{\rm Tr}}

\newcommand{\sfrac}[2]{{\textstyle \frac{#1}{#2}}}
\newcommand{\wh}{\widehat}
\newcommand{\wt}{\widetilde}

\newcommand{\bPi}{\mbox{\boldmath $\Pi$}}

\newcommand{\cE}{{\cal E}}
\newcommand{\cP}{{\cal P}}

\newcommand{\si}{\sigma}

\newcommand{\D}{{\cal D}}
\newcommand{\bD}{{\bf D}}

\begin{document}
\maketitle

\begin{abstract}
The Pauli operator describes the energy of a nonrelativistic quantum
particle with spin $\sfrac{1}{2}$ in a magnetic field and an external
potential. Bounds on the sum of the negative eigenvalues are called
magnetic Lieb-Thirring (MLT) inequalities.  The purpose of this paper
is twofold. First, we prove a new MLT inequality in a simple way.
Second, we give a short summary of our recent proof of a more refined
MLT inequality \cite{ES-IV} and we explain the differences between the
two results and methods.  The main feature of both estimates, compared
to earlier results, is that in the large field regime they grow with
the optimal (first) power of the strength of the magnetic field. As a
byproduct of the method, we also obtain optimal upper bounds on the
pointwise density of zero energy eigenfunctions of the Dirac operator.
\end{abstract}



\medskip\noindent
{\it Running title:} Magnetic Lieb-Thirring inequality for Pauli operator

\section{Introduction}\label{sec:intro}

\subsection{Magnetic Lieb-Thirring inequalities}\label{sec:mlt}

Since the seminal paper of Lieb and Thirring \cite{LT},
Lieb-Thirring inequalities refer to
estimates that bound moments of negative eigenvalues
of Schr\"odinger type operators in terms of the external fields.
They play a fundamental role in various results concerning
localized many-fermion systems. Most notably,
the ground state energy of the many-body Hamiltonian in many cases
is related to the sum of the negative eigenvalues
of an effective one-body Hamiltonian.
Among other useful applications,
Lieb-Thirring inequalities stand behind the most
effective and elegant proofs of stability of
matter. They also serve as a powerful apriori estimate
for the semiclassical analysis of the many-fermion ground state.

We shall not attempt to give an overview of this vast and beautiful
subject, since it is much better to refer the reader
to the
concise review article of Elliott Lieb \cite{L-00}.
We focus instead on the particular case
of {\it magnetic Lieb-Thirring (MLT) inequalities}.
They estimate moments of negative eigenvalues
 $e_1(H)\leq e_2(H)\leq \ldots\leq 0$
 of the Pauli operator
\be
        H:=[\bsigma\cdot (-i\nabla +\bA)]^2 + V\
\label{eq:pauli}
\ee
on $L^2(\bR^3,\bC^2)$
with a vector potential $\bA$,
 magnetic field $\bB:=\nabla\times\bA$ and external potential $V$.
Here $\bsigma\cdot\bv = \si^1v_1 + \si^2v_2 + \si^3v_3$, $\bv\in\bR^3$,
and $\si^1, \si^2, \si^3$ are the Pauli matrices.
Unlike in the nonmagnetic case, where the
{\it optimal form} of the estimates  is well-established
and the remaining main challenge is to find the {\it optimal constants},
the magnetic case is much less understood.
Apart from the case of the constant magnetic field,
which has been settled in \cite{LSY-II} (in two dimensions  \cite{LSY-III}),
so far there is even no conjecture for a general
magnetic Lieb-Thirring inequality that
could be considered optimal in all aspects.

Magnetic fields created in laboratories are usually weak
and with a fair precision they  can be handled perturbatively.
Strong magnetic fields, however,  occur in certain
astrophysical models (e.g. neutron stars, see \cite{LSY-I, LSY-II}
and references therein).
Even a physically weak magnetic field can become effectively
strong in certain problems related to quantum dots \cite{LSY-III}.

Both these physical applications and the spirit of universality
encompassed in the original Lieb-Thirring inequality have led us
to search for magnetic Lieb-Thirring inequalities
on the sum of the negative eigenvalues, that are
optimal as far as the field strength is concerned
in the strong field regime.

The semiclassical formula for the sum of the negative eigenvalues,
$\sum_j |e_j(H)|$, for a constant magnetic field
behaves linearly in the field strength, $|\bB|$ (\cite{LSY-II}).
This fact suggests that $\sum_j |e_j(H)|$
may be bounded by an expression that  grows  with the first power of
$|\bB|$ even for nonconstant magnetic fields and away from
the semiclassical asymptotic regime.
Our goal  is to establish such  MLT
estimates with as few technical assumptions on $\bB$ as possible
and no technical assumptions on $V$.

In this paper we present two such estimates. The simpler bound, Theorem
\ref{thm:main}, is proven in this paper. The more
involved bound, Theorem \ref{thm:long}, is outlined in
 Section \ref{sec:longpaper}, but the details of the
 proof appear elsewhere \cite{ES-IV}.
We point out that the methods behind these two proofs are very different
and they are somewhat complementary, however, we did not succeed in
combining the merits of both.

While both bounds are optimal as far as the potential and the
strength of the magnetic field are concerned, they require additional
technical assumptions on the magnetic field.
These are usually formulated in terms of  {\it variation lengthscales},
and practically they are  regularity assumptions on $\bB$.
This means that supremum norms of derivatives of
the magnetic field appear in the final  Lieb-Thirring inequality.

The difference between the two theorems
is that the more involved bound, Theorem \ref{thm:long},
 involves only local supremum norms. Therefore
it enjoys an important {\it locality property}:
 the estimate is insensitive to the behavior of the magnetic
field far away from the support of $[V]_-$, where $[a]_-:=-\min\{ 0, a\}$
denotes the negative part of $a$.
 The simpler bound, Theorem \ref{thm:main},
involves the global  $C^5$-norm
of the direction of the magnetic field, $\bn : = \bB/|\bB|$.
In particular, irregular behaviour of $\bn$ far away
from the support of $[V]_-$ renders our estimate large
despite that it  should not substantially influence
the negative spectrum.
As a compensation, we need less assumptions on the
regularity of the field strength $|\bB|$ and the proof is
much shorter.

Finally we remark that armed with a MLT inequality that scales
linearly in the field strength, one may prove the naturally expected
semiclassical asymptotics for the sum of the negative
eigenvalues of
$$
        H(h,b):=[\bsigma\cdot (-ih\nabla + b \bA)]^2 + V
$$
uniformly in $b$ as $h\to0$. This requires combining the techniques of
 \cite{ES-II} and \cite{ES-IV} and the details will be published
separately.

\subsection{Short history of magnetic Lieb-Thirring inequalities}

For a constant magnetic field, $\bB\equiv const$, the inequality
\be
        \sum_j |e_j(H)|\leq (const)\Big(
         \int_{\bR^3} |\bB|[V]_-^{3/2} + \int_{\bR^3} [V]_-^{5/2}\Big)
\label{LSYnaive}
\ee
proven in \cite{LSY-II} is optimal, apart from the constant.
It has  seemed to be reasonable to conjecture that (\ref{LSYnaive})
also holds for an arbitrary magnetic field. However,
such a naive generalization fails for two reasons.

Firstly, even when $\bB$ has  constant direction in $\bR^3$
 (\ref{LSYnaive}) can be correct only if $|\bB(x)|$
is replaced by an effective field strength, $B_{\rm eff}(x)$,
 obtained by averaging  $|\bB|$ locally on the magnetic
lengthscale, $|\bB|^{-1/2}$ (\cite{E-1995}).

Secondly, the existence of the celebrated Loss-Yau zero modes
\cite{LY} contradicts  (\ref{LSYnaive}). Indeed, for certain magnetic
fields with  nonconstant direction
the Dirac operator $\D: =\bsigma\cdot(-i\nabla +\bA) $
has a nontrivial $L^2$-kernel. In this case
a small potential perturbation of $\D^2$
shows that  $\sum_j |e_j(H)|$  behaves
as $\int n(x) [V(x)]_- \rd x$, i.e. it is linear in $[V]_-$.
Here $n(x)$ is the density of zero modes, $n(x)= \sum_j |u_j(x)|^2$,
where $\{ u_j\}$ is an orthonormal basis in $\mbox{Ker} \, \D$.
Thus an extra term linear in $[V]_-$ must be added
to (\ref{LSYnaive}) and $n(x)$ has to be estimated.

The problem of the effective field, observed in \cite{E-1995}, was
first succesfully addressed by Sobolev, \cite{Sob-96},  \cite{Sob-97}
and later by Bugliaro et. al. \cite{BFFGS} and Shen
\cite{Sh}. In particular, the $L^2$-norm of
the chosen effective field, $\| B_{\rm eff}\|_2$, is comparable to $\|\bB\|_2$
in \cite{BFFGS}, and the same holds for any $L^p$-norm in
Shen's work. In a very general bound
 proved in \cite{LLS} the first term in (\ref{LSYnaive}) is replaced
with $\| \bB \|_2^{3/2} \| V \|_4$.

In the  works \cite{E-1995}, \cite{Sob-97}, \cite{Sh}, \cite{LLS},
\cite{BFFGS}
the density $n(x)$ is estimated by
a function that behaves quantitatively as $| \bB(x)|^{3/2}$.
In particular, in the strong field regime these estimates
are not sufficient to prove  semiclassical asymptotics
for $H(h,b)$ uniformly in $b$;
they give the asymptotics only up to $b \leq (const.)h^{-1}$
\cite{Sob-98}.

We remark that the bounds in \cite{LLS}
and \cite{BFFGS} have nevertheless been very useful
in the proof of magnetic stability of matter.
In this case the magnetic energy, $\int |\bB|^2$, is
also part of the total energy to be minimized, therefore
even the second moment  of the magnetic field is controlled.
We also remark that if the field has a {\it constant direction},
then no Loss-Yau zero modes exist, $n(x)\equiv 0$.
In this case Lieb-Thirring type bounds that grow linearly with
$|\bB|$ have been proved in \cite{E-1995} and \cite{Sob-96},
\cite{Sob-97}.

Therefore the fundamental difficulty is to understand
the density of Loss-Yau zero modes. It is amusing to note
that it was a substantial endeavour to show
that a zero mode may exist at all \cite{LY},
and that multiple zero modes may also occur \cite{ES-III}.
On the other hand, it seems also quite difficult to give an
upper bound on their number
in terms of the first power of the field strength
(Corollaries \ref{corr:zero} and \ref{cor}).

Since $n(x)$ scales like $(length)^{-3}$ and
$|\bB(x)|$ scales like $(length)^{-2}$, a simple dimension
counting shows that $n(x)$, therefore $\sum_j |e_j|$,
 cannot be estimated in general
by $|\bB(x)|$ or by its smoothed version.
However, if  an extra lengthscale
is introduced, for example certain derivatives of the field are
allowed in the estimate, then it is possible to give
a bound on the eigenvalue sum that grows  slower than $|\bB|^{3/2}$
in the large field regime.
Before the current paper and our recent work \cite{ES-IV}
there were only two results in this direction.

The work \cite{BFG} used a lengthscale
on which $\bB$ changes. The estimate eventually behaves
 like $|\bB|^{17/12}$ in the field strength.
As far as local regularity is concerned,
 only $\bB \in H^1_{loc}$ is required.
However, $n(x)$ is estimated by a quantity that
depends {\it globally} on $\bB(x)$ not just in a neighborhood
of $x$. As a consequence, no locality property holds
for the MLT inequality in \cite{BFG}.

Our earlier work \cite{ES-I} had a different approach to reduce
the power $3/2$ of $|\bB|$
in the estimate of $n(x)$. We introduced two {\it global} lengthscales,
$L$ and $\ell$ respectively,
 to measure the variation scale of the field strength $|\bB|$
and the unit vector
$\bn: =\bB/|\bB|$ that determines the geometry of the field lines.
 This required
somewhat more regularity on $\bB$ than \cite{BFG} and it
also involved the $W^{1,1}$-norm  of $V$.
The estimate grew with the $5/4$-th power of the field strength $|\bB|$
in the large field regime.
For fields with a  nearly constant direction, $\ell\gg 1$,
the bound was actually better, it behaved
like $|\bB|+ |\bB|^{5/4}\ell^{-1/2}$. This indicates
that it is only the variation of $\bn$ and not that of  $|\bB|$
that is responsible for the higher $|\bB|$-power.
The MLT estimate in \cite{ES-I}  did not enjoy
the  locality property either.

Due to the improvement in the $|\bB|$-power
from $3/2$ to $5/4$ in the MLT estimate we could also
prove the semiclassical eigenvalue asymptotics  for $H(h,b)$
in the regime $b \ll h^{-3}$ for potentials in $W^{1,1}$
\cite{ES-II}.
This bound turned out to be sufficient to show
that the Magnetic Thomas-Fermi theory exactly
reproduces the  ground state energy of a large atom with
nuclear charge $Z$ in the semiclassical regime,
i.e. where $b \ll Z^3$, $Z\to\infty$ \cite{ES-II}.
The condition $b\ll Z^3$ is optimal as far as the
semiclassical theory is applicable.

Despite the successful application of the bound in \cite{BFG}
to the stability of matter with quantized electromagnetic field
with an ultraviolet cutoff \cite{BFrG}, and despite that the
MLT inequality given in \cite{ES-I} fully covered the semiclassical
regime of large atoms, it is still important to
establish a uniform Lieb-Thirring type bound with the correct
power in the magnetic field and with no unnecessary assumptions on
$V$.
Such bound will likely be the key
to generalize the analysis of the super-strong field regime
of \cite{LSY-I} to non-homogeneous magnetic fields.

\bigskip

\section{Lieb-Thirring inequality without locality property}\label{sec:global}
\setcounter{equation}{0}

We consider the
 three dimensional Pauli operator,
$ H= \D^2+V$,
with a differentiable magnetic field $\bB=\nabla\times\bA$.
The operator
\be
\D:=\bsigma\cdot (-i\nabla +\bA)
\label{eq:origdi}
\ee
is called the Dirac operator
with vector potential $\bA$ and
magnetic field $\bB$.
We make two global assumptions:

\medskip

{\it Assumption 1.} $\bB(x)\neq 0$  for all $x\in\bR^3$, i.e. the
 unit vectorfield $\bn:= \bB/|\bB|$ is well defined.

\medskip

{\it Assumption 2.} The vectorfield $\bn$
satisfies the following global regularity condition
\be
      L_\bn^{-1}:=\sum_{\gamma =1}^5 \| \nabla^\gamma \bn\|_\infty^{1/\gamma}<
      \infty\; ,
\label{eq:Ln}
\ee
where $L_\bn$ is called the {\it global variation lengthscale} of $\bn$.

\medskip

For any $L>0$, $x\in\bR^3$ we also define
$$
        B^*_L (x):=\sup\{ |\bB(y)|\; : \; |y-x|\leq L\}\;
        + L \cdot \sup\{ |\nabla\bB(y)|\; : \; |y-x|\leq L\}\; .
$$

\begin{theorem}[Lieb-Thirring inequality without a locality property]
\label{thm:main} For any $0<L\leq L_\bn$,
the sum of the negative eigenvalues, $e_1(H)\leq e_2(H)\leq \ldots \leq 0$,
 of $H$ satisfies
\be
       \sum_j |e_j(H)|
        \leq c \Big( L^{-1} \int (B^*_L+L^{-2}) [V]_-
      + \int  B^*_L [V]_-^{3/2} + \int [V]_-^{5/2}\Big)
\label{eq:LT}
\ee
with a universal constant $c$.
\end{theorem}

Let $\Pi_{\D^2\leq \lambda}$ denote the spectral projection
onto energy levels below $\lambda$ in the spectrum of $\D^2$.
Along the proof of Theorem \ref{thm:main} we obtain the following bound
on the density of the
low lying states of $\D^2$:

\begin{corollary}[Density of low lying states]\label{corr:zero}
 For any $0<L\leq L_\bn$
and $x\in\bR^3$
\be
        \Pi_{\D^2\leq \lambda}(x,x)\leq c
        \lambda_*^{1/2}\max\{ B_L^*(x), \lambda_*\} \; ,
        \quad \mbox{with} \quad \lambda_*:=\max\{ \lambda, L^{-2}\},
\label{eq:corr}
\ee
in particular the local density of zero modes of $\D^2$
grows with at most
 the first power of the field strength.
\end{corollary}

\medskip

{\it Convention.} We use the letter $c$ to denote various positive
universal constants whose exact value may change from line to line.

\section{Lieb-Thirring inequality with a locality
property}\label{sec:longpaper}
\setcounter{equation}{0}

In this Section we give an outline of a more refined magnetic
Lieb-Thirring inequality. The detailed proof appears elsewhere
\cite{ES-IV}.

We assume that $\bB \in C^4(\bR^3,\bR^3)$,
and define three basic lengthscales of $\bB$.
The Pauli operator will be localized on these
lengthscales.
Let
$\bn:=\bB/|\bB|$ be the unit vectorfield in the direction
of the magnetic field at all points where $\bB$ does not vanish.
For any $L>0$ and $x\in \bR^3$ we define
\be
    B_L(x): = \sup\{ |\bB(y)|\; : \; |x-y|\leq L \}
\label{eq:BL} \; ,
\ee
and
\be
    b_L(x): =  \inf\{ |\bB(y)|\; : \; |x-y|\leq L \}
\label{eq:bL}
\ee
to be the supremum and the infimum
of the magnetic field strength
on the ball of radius $L$ about $x$.

\begin{definition}[Lengthscales of a magnetic field]\label{def:mag}
 The
{\bf magnetic  lengthscale}  of $\bB$ is defined as
$$
       L_m(x):= \sup\{ L>0 \; : \; B_L(x) \leq L^{-2}\} \; .
$$
The {\bf variation lengthscale} of $\bB$ at $x$ is given by
$$
       L_v(x): = \min \{ L_s(x), L_n(x) \}\; ,
$$
where
$$
       L_s(x): = \sup\Big\{ L>0 \; : \; L^\gamma \sup\Big\{ \Big|\nabla^\gamma
       |\bB(y)| \;\Big| \; : \; |x-y|\leq L\Big\} \leq b_L(x),
       \; \gamma=1,2,3,4 \Big\} \;
$$
$$
       L_n(x): = \sup\Big\{ L>0 \; : \; L^\gamma \sup\{ |\nabla^\gamma
       \bn(y)|\; : \; |x-y|\leq L, \; \bB(y)\neq 0\}
       \leq 1, \; \gamma=1,2,3,4 \Big\} \;
$$
(with the convention that $\sup\emptyset =-\infty$).
Finally we set
\be
       L_c(x): = \max \{ L_m(x), L_v(x) \} \; .
\label{def:Lc}
\ee
\end{definition}

A magnetic field $\bB:\bR^3\to\bR^3$ determines three
local lengthscales. The magnetic lengthscale, $L_m$, is
comparable with
$|\bB|^{-1/2}$. The lengthscale $L_s$ determines the
scale on which the strength of the field varies,
i.e., it is the variation scale of $\log |\bB|$.
The field line structure, determined by $\bn$, varies on
the scale of $L_n$. The variation lengthscale $L_v$
is the smaller of these last two scales, i.e.,
it is the scale of variation of the vectorfield $\bB$.

For weak magnetic fields the magnetic effects can be neglected
in our final eigenvalue estimate, so the variational
lengthscale becomes irrelevant. This idea is reflected
in the definition of $L_c$; we will not need to localize
on scales shorter than the magnetic scale $L_m$.

The following theorem is the main result in \cite{ES-IV}.

\begin{theorem}[Lieb-Thirring inequality with a locality property]
\label{thm:long}
 We assume that the magnetic field  $\bB=\nabla\times\bA$
is in $C^4(\bR^3, \bR^3)$.
Then the sum of the negative eigenvalues of
 $H=  [\bsigma\cdot (\bp+\bA)]^2 + V$ satisfies
\be
         \sum_j |e_j| \leq c \int [V]_-^{5/2}
       + c\int |\bB|[V]_-^{3/2} + c\int (|\bB|+L_c^{-2})L_c^{-1} [V]_-\; .
\label{eq:main}
\ee
\end{theorem}

\begin{corollary}[Density of zero modes with a locality property]\label{cor}
Given a magnetic field $\bB\in C^4(\bR^3, \bR^3)$,
the density of zero modes of the free Dirac operator
$\D$ with magnetic field $\bB$ satisfies
\be
      n(x):= \sum_j |u_j(x)|^2 \leq c(|\bB(x)|+L_c^{-2}(x))L_c^{-1}(x)\; ,
\label{eq:nest}
\ee
where $\{u_j\}$ is an orthonormal basis
in the kernel of $\D=\bsigma\cdot(-i\nabla+\bA)$.
\end{corollary}

{\it Remarks.} (i)
 The bound (\ref{eq:nest}) is optimal as far as the strength
of the field $|\bB|$ is concerned. This fact follows from
the construction of Dirac operators with kernels of high multiplicity
following the method of \cite{ES-III}.

(ii) Notice that the Lieb-Thirring inequality of \cite{LSY-II} for a
 {\it constant} field is recovered  in Theorem \ref{thm:long}.
However, the uniform Lieb-Thirring bound for a {\it constant direction}
field, \cite{Sob-97}, \cite{ES-I}, does not directly follow from
our main theorem as it is stated. Firstly, (\ref{eq:main})
contains a term linear in $V$ that is unnecessary for a constant
direction field. Secondly, we assume high regularity on
$\bB$.  This regularity is needed only to construct
the appropriate curvilinear cylindrical localization, which
is unnecessary for a field with  constant direction.

(iii) It can be shown that $L_c(x)$ enjoys a locality property;
for any $\delta>0$ the inverse lengthscale $L_c^{-1}(x)$ is
bounded by a $\delta$-dependent function  of the local $C^4$-norm
of $\bn$ and $\log |\bB|$ in a $\delta$-neighborhood of $x$.

\bigskip

Before we turn to the proof of our new result, Theorem \ref{thm:main},
we explain the key difference between the two  proofs
that is somewhat hidden behind technicalities.

The linear power of $|\bB|$ in the estimate reflects
the basic fact that  the space with a magnetic field
cannot be considered isotropic: the quantum motion
parallel with the magnetic field behaves differently
than the transversal one.
The magnetic field affects only the transversal motion;
this is why the eigenvalue sum scales in the same (linear)
power of $|\bB|$ both in $d=2$ and 3 dimensions.
The motion in the third direction affects only the
powers of the potential, as it is customary in the
nonmagnetic Lieb-Thirring inequalities.

All MLT estimates that yield $|\bB|^{3/2}$
behaviour  neglect this geometric fact
by simply comparing the magnetic problem
with a nonmagnetic one, usually via
a diamagnetic inequality that loses
the anisotropic feature of the problem.
The typical estimate is of the form
\be
        \D^2 \ge b^{-1} \D^2 = b^{-1}
        ( \bD^2 + \bsigma\cdot\bB)
\label{eq:li}
\ee
where $\bD:=\bp +\bA$ is the spinless magnetic momentum
and $b:= \|\bB\|\gg 1$ denote some (local) norm of $\bB$.
The Pauli kinetic energy is scaled down so that the dangereous
$\bsigma\cdot\bB$ becomes bounded uniformly in $b$. The magnetic Laplacian
$\bD^2$  then can be controlled by the nonmagnetic Laplacian,
$-\Delta$, but the
factor  $b^{-1}$ now affects all three coordinates,
yielding a scaling of $b^{3/2}$.
The key to the improvement is to separate
the motion in parallel and in the transversal directions
and use a crude estimate similar to
(\ref{eq:li}) only in the two-dimensional transversal
kinetic energy.

Since the direction of the magnetic field $\bn$
varies, it is in general impossible to introduce
global orthogonal coordinates parallel and transversal to
$\bn$. The more conventional approach is the one presented
in \cite{ES-IV}: we introduce appropriate coordinates
locally and approximate the magnetic field by another
one that is constant in these coordinates. For constant
magnetic field the separation is then trivial. The main
technical challenge is to keep all approximation and
localization errors bounded uniformly in the field strength $b$.
This requires many steps: most importantly we have to change the metric
by a conformal factor to make the field strength constant along the
field lines; we have to localize the
problem onto the physically correct domains, which
are elongated curvilinear cylinders of width $b^{-1/2}$
along the field lines and we have to construct appropriate coordinate
systems along each field line.
 The construction steps require
tools from the spinor geometry behind the Dirac operators.
The localization onto narrow cylinders is done via
the magnetic localization formula from \cite{ES-II}.
All these steps require fairly high regularity on the
magnetic field.

The more direct  approach is the one behind Theorem \ref{thm:main}.
This method bypasses all the complications
with the constant field approximation, the cylindrical
construction and the change of metric. It uses the simple observation
that the energy in the parallel direction
can be extracted from $\D^2$ by a simple energy argument
(see (\ref{eq:third}) in Lemma \ref{lemma:lower}):
\be
        \D^2 \ge c D_{\bn}^2 - (error) \; ,
\label{eq:dbn}
\ee
where $D_\bn: =-i\partial_\bn -\sfrac{i}{2}\mbox{div}\, \bn$,  and
 the error terms are of lower order in momentum.
It turns out to be necessary to consider higher powers of
the resolvent $(\D^2 + \lambda)^{-1}$
to control the ultraviolet regime (here $\lambda>0$ is a constant
controlled by  the geometry of the field lines),
therefore it is necessary to estimate higher powers of
$\D^2$ as well, the key observation is that
after having extracted $D^2_\bn$, higher powers of
the crude diamagnetic estimate (\ref{eq:li}) are sufficient. Although
(\ref{eq:li}) cannot be squared directly,
after several commutators it is possible to show
that (see (\ref{eq:cube}))
\be
        \D^{2k} \ge [\delta\bD^2]^k - (error) \; ,
\label{eq:tdk}
\ee
where $\delta:=\min\{ \lambda b^{-1}, 1\}$.

Therefore the basic estimate for the density of
the low lying states is
$$
        \Pi_{\D^2\leq \lambda} \leq \frac{8\lambda^3}{(\D^2+\lambda)^3}
        \leq \frac{c\lambda^3}{ \D^6+\lambda^2 D_\bn^2+\lambda^3}
        \leq \frac{c\lambda^3}{  (\delta\bD^2)^3+\lambda^2 D_\bn^2+\lambda^3}\; .
$$
Then we can essentially use the diamagnetic inequality
(although first we have to ensure that the
two operators in the denominator commute, see Lemma \ref{lemma:comm})
and the diagonal term is
estimated as
$$
        \Pi_{\D^2\leq \lambda}(x,x)
        \leq \frac{c\lambda^3}{ \delta^{3} (-\Delta)^3+\lambda^2 \partial^*_\bn
        \partial_\bn+\lambda^3}
        \, (x,x) \; .
$$
It is then easy to change this quantity into Euclidean coordinates
and compute
$$
        \int_{\bR^3}
         \frac{\lambda^3}{ \delta^{3} ({\bf p}^2)^3+\lambda^2 p_3^2+\lambda^3}\; \rd
        {\bf p}
        \leq c\lambda^{1/2}(b+\lambda)
$$
which scales linearly with $b$.

Apart from several technicalities, there is an apparently innocent
but fundamental complication in both approaches.
Both proofs uses the concept of the local magnetic field
strength, $b$, that is given by some local norm of $\bB$.
Therefore at the beginning one needs  to localize
the original operator $\D^2$ onto domains of size
independent of $b$ to establish this apriori localization scale.
We  choose domains of size $L_\bn$
where the field lines do not vary substantially.

Localizing higher powers of $\D^2$ even onto the scale
$L_{\bn}$ is difficult because it does not follow simply from energy
considerations. If the localization length $L_\bn$ is a uniform constant
then it can be done using the spectral theorem
(see (\ref{eq:spe})). This is the
method in Theorem \ref{thm:main}, but it forces one to
introduce the  same localization error everywhere in space,
even in regimes, where the field is very regular.

If one wants to include a variable $L_\bn$, as
in Theorem \ref{thm:long}, to ensure the locality property,
then one needs a more powerful apriori localization.
Typically it is not hard to localize resolvents of second order
elliptic operators
onto cubes of size $\ell$ at the expense of an error $\ell^{-2}$.
However localizing the square (or higher powers) of the
resolvent requires off-diagonal estimates on the resolvent
kernel (see Proposition 7.1 in \cite{ES-IV}).
 While these are typically easily available for
scalar elliptic operators without spin,
 we {\it do not know any apriori
off-diagonal control on the resolvent of $\D^2$}.
If the original Pauli operator is estimated by a constant field
Pauli operator, then {\it aposteriori} we can extract
off-diagonal estimates, and this idea is used, though
very implicitly, in \cite{ES-IV}. But without comparison
with the constant field problem, we do not have
off-diagonal control. This is the main reason why we are unable
to extend the elegant and short method of Theorem \ref{thm:main}
to give any locality properties.

\section{Organization of the proof of Theorem \ref{thm:main}}\label{sec:org}
\setcounter{equation}{0}

We can assume that the potential in (\ref{eq:pauli})
is nonpositive, so for simplicity we can
consider $H=\D^2-V$ with $V\ge 0$.
We start with the Birman-Schwinger  principle
\be
   |\Tr (\D^2-V)_-| = 2\int_0^\infty
   n\Big( |V-E|_-^{1/2} \frac{1}{\D^2+E}|V-E|_-^{1/2} \; , 1\Big)\rd E \; ,
\label{eq:BS}
\ee
where $n(X ; c)$ denotes the number of eigenvalues
of the positive self-adjoint operator $X$ above  level $c$.

We  estimate $\D^2$ from below by the magnetic Laplacian,
then use the diamagnetic inequality to estimate its resolvent
kernel by that of the free Laplacian. These estimates
involve error terms that cannot be controlled by $E$ in
the resolvent $(\D^2+E)^{-1}$ when $E$ is small. As a first step,
we insert a positive constant $P$ in the resolvent using the
inequality
\be
     \frac{1}{\D^2 + E} \leq \frac{2}{\D^2 +P+ E} + \frac{8}{E}
      \frac{P^3}{(\D^2 +P)^3}
\label{eq:spe}
\ee
that follows from the spectral theorem and the corresponding arithmetic
inequality. $P$ is chosen  as
$$
     P: = \e^{-4}L^{-2}\; ,
$$
where $\e$ denotes
a sufficiently small universal constant.

{F}rom (\ref{eq:spe})
and the general bound
 $n(X_1 + X_2; c_1 + c_2) \leq n(X_1, c_1) + n(X_2, c_2)$, we obtain
\be
   |\Tr (\D^2-V)_-|  \leq c[ (I) + (II)]
\label{eq:split}
\ee
with
\be
        (I): = \int_0^\infty
          n\Big( |V-E|_-^{1/2} \frac{1}{\D^2+P+E}|V-E|_-^{1/2} \; ,
        \sfrac{1}{4}\Big)\rd E \; ;
\label{eq:Idef}
\ee
\be
       (II):= \int_0^\infty
          n\Big( |V-E|_-^{1/2} \frac{P^3}{(\D^2+P)^3}|V-E|_-^{1/2} \; ,
        \sfrac{1}{16} E\Big)\rd E \; .
\label{eq:IIdef}
\ee
These two terms will be called {\it positive energy regime} and
{\it zero mode regime}, respectively (see \cite{ES-IV}).

The proof of Theorem \ref{thm:main} then follows from the
 two estimates below:

\begin{proposition}\label{prop:main}
For sufficiently small $\e$
\be
        (I) \leq c \int B_L^*(x) V^{3/2}(x)\rd x
        + c\int V^{5/2}
\label{eq:Ires}
\ee
and
\be
        (II)\leq c P^{1/2}\int \max\{ B_L^*(x), P\} V(x)\rd x \; .
\label{eq:IIres}
\ee
\end{proposition}

To obtain these estimates, we will have
 to bound the resolvent of $\D^2$
and the diagonal kernel of the cube of the resolvent of $\D^2$
(see (\ref{eq:IItrace})).
In both cases we  localize the operator $\D^2$ onto
cubes of size of order $L^{-1}$.

In Section \ref{sec:loc} we define localization
functions and new Dirac operators which coincide with $\D$
locally but have a uniformly bounded magnetic field everywhere.
In Section \ref{sec:bdd} we prove various estimates
on a Dirac operator with a bounded magnetic field and we compare
it with an auxiliary Laplacian-type elliptic operator
that is diagonal in the spin space.
Finally, in Sections \ref{sec:loccube} and
\ref{sec:res} we complete the estimates (\ref{eq:IIres})
and (\ref{eq:Ires}), respectively.

\section{Localization}\label{sec:loc}
\setcounter{equation}{0}

Consider the cubic lattice $\Lambda=\e L \bZ^3$. For each $k\in \Lambda$,
let $\Omega_k\subset \Omega_k'\subset \Omega_k^*\subset
 \Omega_k^\#$ be the balls with center $k$ and with
radius $\e L$, $2\e L$, $3\e L$ and
$4\e L$, respectively.
Let $\{\theta_k\}_{k\in \Lambda}$ be a partition of unity
such that $0\leq \theta_k\leq 1$,
\be
        \sum_{k\in\Lambda}
        \theta_k^2\equiv 1, \qquad \mbox{supp} \; \theta_k\subset \Omega_k,
        \qquad \|\nabla^\gamma\theta_k\|_\infty\leq c (\e L)^{-\gamma},
        \quad 1\leq \gamma\leq 5\; .
\label{eq:theta}
\ee
We also choose further localization
functions $\eta_k$, $\omega_k$ and $\chi_k$ with
$\theta_k\leq \eta_k \leq \omega_k
\leq \chi_k \leq 1$, $k\in \Lambda$, such that
\be
        \mbox{supp} \; \eta_k\subset \Omega_k', \qquad \eta_k\equiv 1
        \quad \mbox{on} \;\; \Omega_k,
        \qquad  \|\nabla^\gamma\eta_k\|_\infty\leq c(\e L)^{-\gamma},
        \quad 1\leq \gamma\leq 5\; ;
\label{eq:eta}
\ee
\be
        \mbox{supp} \; \omega_k\subset \Omega_k^*, \qquad \omega_k\equiv 1
        \quad \mbox{on} \;\; \Omega_k',
        \qquad \|\nabla^\gamma\omega_k\|_\infty\leq c(\e L)^{-\gamma},
        \quad 1\leq \gamma\leq 5\; ;
\label{eq:omega}
\ee
\be
        \mbox{supp} \; \chi_k\subset \Omega_k^\#, \qquad \chi_k\equiv 1
        \quad \mbox{on} \;\; \Omega_k^*,
        \qquad \|\nabla^\gamma\chi_k\|_\infty\leq c(\e L)^{-\gamma},
        \quad 1\leq \gamma\leq 5\; .
\label{eq:chi}
\ee
We  note that
\be
        \sum_{k\in\Lambda} \eta_k \leq C, \qquad
        \sum_{k\in\Lambda} \omega_k \leq C,
         \qquad         \sum_{k\in\Lambda} \chi_k \leq C
\label{eq:fin}
\ee
for some universal constant $C$
by the finite overlap property of the balls.

Let $b_k:=\sup_{\Omega_k^\#}(|\bB|+L|\nabla \bB|)$,
then we have
\be
        |\bB|+L|\nabla \bB|
        \leq b_k \leq B^*_L \quad \mbox{on} \;\; \mbox{supp} (\chi_k) \; ,
\label{eq:bkb}
\ee
if $\e$ is sufficiently small.
\medskip

The following lemma defines a new Dirac operator
for each cube $\Omega_k$ that coincides with $\D$ on $\Omega_k^*$ but
has a uniformly bounded magnetic field. The proof is given in
Section \ref{sec:extproof}.

\begin{lemma}\label{lemma:ext}
For each $k\in \Lambda$ there exists a nowhere vanishing
 magnetic field $\bB_k$ such that $\bB_k \equiv \bB$ on $\Omega_k^*$,
$\bB_k\equiv (const)$ on $\bR^3\setminus \Omega_k^\#$,
\be
        \|\bB_k\|_\infty + L\|\nabla\bB_k\|_\infty\leq cb_k,
\label{eq:bkbound}
\ee
 and $\bn_k:=\bB_k/|\bB_k|$ satisfies
the bound
\be
        \sum_{\gamma=1}^5(\e L)^\gamma
        \|\nabla^\gamma \bn_k\|_\infty
        \leq c \e\; .
\label{eq:Lnk}
\ee
Moreover, there exists a vector potential $\bA_k$, $\nabla\times \bA_k=\bB_k$
such that $\bA_k\equiv \bA$ on $\Omega_k^*$. The Dirac operator
$\D_k: = \bsigma\cdot (\bp + \bA_k)$ with magnetic field $\bB_k$ therefore
satisfies
\be
        \D_k\eta_k = \D\eta_k\; .
\label{eq:deqd}
\ee
\end{lemma}

The localization is performed by the following ``Pull-up''
formula whose proof was given in \cite{BFFGS} and was also recalled in
 Proposition 6.1. in \cite{ES-IV}.

\begin{proposition}[Pull-up formula]\label{prop:pullup}
Let $I$ be a countable index set
and let $g_i$, $i\in I$,
 be a family of nonnegative functions on $\bR^3$ such
that $\sum_{i\in I} g_i^2(x) <\infty$ for every $x\in \bR^3$.
Let $A_i$, $i\in I$, be a family of positive invertible self-adjoint operators
on $L^2(\bR^3; \bC^2)$. Then
\be
    \Big(\sum_{i\in I}g^2_i \Big)
    {1\over \sum_{i\in I} g_i A_ig_i}
     \Big(\sum_{i\in I}g^2_i \Big) \leq \sum_{i\in I}
     g_i \; {1\over A_i} \; g_i\;.
\label{eq:pullup}
\ee
\end{proposition}

\bigskip

The resolvent $(\D^2+P+E)^{-1}$ can be easily localized
using the partition of unity $\{\theta_k\}$.
$$
        \D^2 = \sum_k\D\theta_k^2\D \ge \frac{1}{2} \sum_k
        \theta_k \D^2\theta_k
         -\sum_k|\nabla \theta_k|^2
        \ge \frac{1}{2} \sum_k
        \theta_k \D_k^2\theta_k - c\e^{-2} L^{-2}
$$
using (\ref{eq:deqd}), the  finite overlap property of
the supports of $\theta_k$'s,
and the estimate on their derivatives (\ref{eq:theta}).
For sufficiently small $\e$ the error can be absorbed into $P$, so
$$
        \D^2+P+E \ge \frac{1}{2} \sum_k
        \theta_k (\D_k^2+P+E)\theta_k \;.
$$
Using the ``Pull-up'' formula (\ref{eq:pullup}),
we obtain
\be
        \frac{1}{ \D^2+P+E } \leq
        \frac{2}{  \sum_k
        \theta_k (\D_k^2+P+E)\theta_k}
        \leq 2\sum_k \theta_k \; \frac{1}{ \D^2_k+P+E } \; \theta_k \; .
\label{eq:pullup1}
\ee

Similar localization does not hold for higher powers
of the resolvent as it was remarked in Section 7 of \cite{ES-IV}.
However, we can localize the second term in (\ref{eq:spe})
estimating it by an auxiliarty elliptic operator that is
independent of the spin coordinates. The construction is given
in Section \ref{sec:bdd}.

\section{Analysis of bounded magnetic fields}\label{sec:bdd}
\setcounter{equation}{0}

Throughout this section we assume that a vector potential $\bA$ is
given so that for some positive constants
$L, \e$ and $b$,
the nowhere vanishing magnetic field $\bB=\nabla\times \bA$
of the Dirac operator $\D = \bsigma\cdot(\bp+\bA)$
satisfies the following conditions:
\be
      \bB \equiv \mbox{const} \qquad \mbox{outside of}  \;\;\Omega^\#,
\label{eq:const}
\ee
where $\Omega^\#$ is a ball of size $O(\e L)$;
\be
   0< |\bB| + L|\nabla \bB|\leq b
\label{eq:bb}
\ee
and
\be
        \sum_{\gamma=1}^5 (\e L)^\gamma \|\nabla^\gamma \bn\|_\infty\leq
        c\e \; ,
\label{eq:unifn}
\ee
where $\bn:= \bB/|\bB|$.
The plane $\cP$ through the center of $\Omega^\#$ and orthogonal
to the magnetic field at the center is called the supporting plane.

The results of this section will be applied to the Dirac
operators $\D_k$ with a magnetic field $\bB_k$
constructed in Lemma \ref{lemma:ext}, but we do not carry
the index $k$ inside this section.

\bigskip

We choose a global, positively oriented
 orthonormal basis $\{ e_1, e_2, e_3\}$ in $\bR^3$
whose third vector is $e_3:=\bn$,
and such that all Christoffel symbols $\Gamma_{kj}^m = (\nabla_{e_j}e_k, e_m)$
vanish outside the
ball twice bigger than $\Omega^\#$, and they satisfy the estimate
\be
        \sum_{\gamma=0}^4(\e L)^\gamma\|\nabla^\gamma \Gamma_{kj}^m\|_\infty\leq cL^{-1} \; .
\label{eq:ch}
\ee
That such and othonormal basis exists follows easily from
(\ref{eq:unifn}), e.g., by a Gram-Schmidt procedure starting from
$\bn, \widetilde{e}_1,\widetilde{e}_2$, where
$\widetilde{e}_1,\widetilde{e}_2$ are fixed vectors.
 We will work on the trivial spinorbundle over $\bR^3$, i.e.,
in the $L^2(\bR^3)\otimes \bC^2$ space. We briefly recall the appropriate
formalism from Section 9 of \cite{ES-IV}.

For any 1-form $\alpha$ we define the  spinor connection
$$
        \nabla_X^\a :=\partial_X + i\a(X) +\sfrac{i}{2}
        \bsigma\cdot \bomega(X) \; ,
$$
where
$$
        \bomega(X):=\Big( (\nabla_X e_3, e_2),
(\nabla_X e_1, e_3), (\nabla_X e_2, e_1)\Big)\; ,
$$
and let
$$
        \Pi_j^\a : = -i\nabla_{e_j}^\a , \qquad j=1,2,3 \; .
$$
For simplicity we will use  $\nabla_j: = \nabla_{e_j}$
and $\partial_j = \partial_{e_j}$ and
 we usually omit the superscript $\a$ from the notation.
 We recall that the operator
\be
        \wh D_X:=-i\partial_X -\sfrac{i}{2} \, \mbox{div} \; X
\label{def:wtD}
\ee
 is self-adjoint on $L^2(\bR^3)$ for any $C^1$ bounded
vectorfield $X$, therefore
the operator
\be
         D_X : = -i\partial_X -\sfrac{i}{2} \mbox{div}\; X
        + \alpha(X)
\label{def:D}
\ee
is self-adjoint. The hat refers to the operators without magnetic field,
$\a \equiv 0$.

These operators are originally defined on
$L^2(\bR^3)$, but with a little abuse of notations
they can also be viewed as acting on
$L^2(\bR^3)\otimes\bC^2$  as $D_X\otimes I_2$, where $I_2$ is
the identity operator on the spin space. We will not distinguish
between these two operators in the notations.
We set $D_j: = D_{e_j}$ for simplicity, then
\be
        \Pi_j = D_j + \sfrac{i}{2}\mbox{div} \; e_j + \sfrac{1}{2}
        \bsigma\cdot \bomega
        (e_j) \; .
\label{eq:pid}
\ee
We compute the
 commutators of the $D_j$'s:
\be
    [D_j, D_k] = (-i) (D_{[e_j, e_k]} + \beta_{jk})
\label{eq:dcomm}
\ee
where $\beta:=\rd \a$ and
$\beta_{jk}:= \beta(e_j, e_k)$.
In the computation
we used that  the divergence part vanishes since
\be
    X\mbox{div}\, Y - Y\mbox{div}\, X - \mbox{div}\,[X,Y]=0
\label{div}
\ee
for any vectorfields. The relation (\ref{div}) is easy to check on
 coordinate vectorfields and it is tensorial.

Let $ Q_{jk}^m: = \Gamma_{kj}^m - \Gamma_{jk}^m$, then
$D_{[e_j, e_k]} =\sfrac{1}{2}(Q_{jk}^mD_m +D_m Q_{jk}^m)$,
where we used the summation convention for repeated indices.
{F}rom (\ref{eq:dcomm})
\be
    [D_j, D_k] =-\sfrac{i}{2}( Q_{jk}^mD_m +D_m Q_{jk}^m) -i\beta_{jk}=
          (-i) \Big( Q_{jk}^mD_m -\frac{i}{2}(\partial_m Q_{jk}^m)
    + \beta_{jk}
    \Big) \; .
\label{eq:commD}
\ee

We recall from Section 9 of  \cite{ES-IV} that $\nabla_X^\a$
is a spinor connection with a magnetic 2-form $\beta:=\rd\a$.
If we choose the vector potential $\alpha (X): = \bA \cdot X$,
then $\star\beta$ is the 1-form corresponding to the vectorfield $\bB$,
where $\star$ denotes the Hodge dual with respect to the standard
Euclidean metric. With these notations  the
Dirac operator $\D = \bsigma\cdot(\bp+\bA)$ in Euclidean coordinates
 is $SU(2)$-gauge equivalent to
\be
        \wt\D := \bsigma\cdot \bPi \; ,
\label{def:hatd}
\ee
i.e.,
\be
        \D = U \wt D U^*
\label{eq:unit}
\ee
for some $U: \bR^3\mapsto SU(2)$. Tilde always refers to Dirac operators
written in a basis where the third basis element is parallel with
the magnetic field. In the chosen basis
the magnetic 2-form satisfies $\beta(e_1, e_2) = B$, where $B:=|\bB|$
is the field strength, and all other $\beta(e_j, e_k) =0$.

The Lichnerowicz formula is given by
\be
        \wt\D^2 = \bPi^*\cdot \bPi + \frac{1}{4} R_0 + \sigma(\star \beta)\; ,
\label{eq:lich}
\ee
where $R_0$ is the scalar curvature (which vanishes in our case)
and for any one-form $\lambda=\sum_j
\lambda_je^j$
we define $\sigma(\lambda)=\sum_j \si^j\lambda_j$.

\medskip

Let $Z$ be a symmetric positive definite 3 by 3 matrix valued
function on $\bR^3$ that satisfies
\be
        \| Z- I \| + \sum_{\gamma=1}^3 (\e L)^\gamma \| \nabla^\gamma Z \|
         \leq c\e\; ,
\label{eq:ZI}
\ee
and let $h$ be a function with
\be
          \sum_{\gamma=0}^2 (\e L)^\gamma
         \|\nabla^\gamma h\|_\infty \leq c\e^{-1}L^{-2}\; .
\label{eq:hest}
\ee

We define
\be
        T:=T_{Z,h}:= \bD\cdot Z \bD + h, \qquad
        \wh T:=\wh T_{Z,h}:= \wh \bD\cdot Z \wh \bD + h
\label{def:tzh}
\ee
on $L^2(\bR^3)$, but we will also denote simply by $T$, $\wh T$
the operators $T\otimes I_2$ and $\wh T\otimes I_2$ acting on
$L^2(\bR^3)\otimes \bC^2$.
We will choose $Z$ and $h$ according to the following lemma
whose proof is given in Section \ref{sec:comm}.

\begin{lemma}\label{lemma:comm}
 Let the magnetic field $\bB=\nabla\times \bA$ satisfy (\ref{eq:const}),
 (\ref{eq:bb}), and (\ref{eq:unifn}).
Then for sufficiently small $\e$
there exists a 3 by 3 symmetric matrix valued function $Z$ and
a real function $h$ satisfying (\ref{eq:ZI}) and (\ref{eq:hest}),
such that the commutator of $T= \bD\cdot Z \bD + h$ and $D_\bn$
vanishes,
\be
        [ T, D_\bn ] = 0 \; .
\label{eq:dzdcomm}
\ee
\end{lemma}

\medskip

Armed with these notations, we state the following lower bounds
on the Pauli operator. The proof is given in Section
\ref{sec:lowproof}.

\begin{lemma}\label{lemma:lower}
Let the magnetic field $\bB$ of the Dirac operator $\wt\D=\bsigma\cdot \bPi$
satisfy (\ref{eq:const}), (\ref{eq:bb}), and (\ref{eq:unifn}).
Let $0\leq \theta\leq
\eta\leq \omega\leq \chi\leq 1$ be real functions on $\bR^3$
with $|\nabla^\gamma\theta|, |\nabla^\gamma\eta|,|\nabla^\gamma\omega|,
|\nabla^\gamma\chi|\leq c(\e L)^{-\gamma}$,
$1\leq \gamma\le 5$, and $\eta\equiv 1$ on $\mbox{supp}(\theta)$,
$\omega\equiv 1$ on $\mbox{supp}(\eta)$ and
$\chi\equiv 1$ on $\mbox{supp}(\omega)$.
Let $P:=\e^{-4}L^{-2}$, $\lambda\ge P$ and $\delta:=
\min \{ \lambda b^{-1}, 1\} $.
Then for sufficiently small $\e$ there exists a positive universal
constant $c$ such that
\be
        \wt\D^2 \ge c \, D_\bn^2 - \e^2 P \;
\label{eq:third}
\ee
and
\be
        \wt\D^3 \theta^2 \wt\D^3 + \lambda\wt\D^2\eta^2 \wt\D^2
        + \lambda^2\wt\D\omega^2 \wt\D+\lambda^3\chi^2
          \ge c \, \theta (\delta  T)^3 \theta \; ,
\label{eq:cube}
\ee
where $T=T_{Z,h}$ is defined in (\ref{def:tzh}) with the
bounds (\ref{eq:ZI}), (\ref{eq:hest}).
\end{lemma}

Finally, we estimate the diagonal of the resolvent kernel
of the sum of the lower bounds obtained  in Lemma \ref{lemma:lower}.
The proof is given in \ref{sec:Tdiagproof}.

\begin{lemma}\label{lemma:Tdiag}
Let the magnetic field $\bB=\nabla\times\bA$
satisfy (\ref{eq:const}), (\ref{eq:bb}), and (\ref{eq:unifn}),
and let $\lambda\ge \e^{-4} L^{-2}$, $\delta:=
\min \{ \lambda b^{-1}, 1\} $.
Then for sufficiently small $\e$
\be
 \frac{1}{ (\delta T )^3 +
         \lambda^2 D_{\bn}^2 + \lambda^3}(x,x)\leq c \lambda^{1/2}\max\{ b,
        \lambda\} \, .
\label{eq:diag}
\ee
\end{lemma}

\section{Proof of the zero mode regime}
\label{sec:loccube}
\setcounter{equation}{0}

In this section we prove a Lemma that immediately
implies (\ref{eq:IIres}).
 The main tool is a localization
of the third power of the resolvent.
For later purposes,
we will need the following, somewhat more
general estimate:
\begin{lemma}\label{lemma:cubeloc} Assume that the magnetic field
$\bB$ of a Dirac operator $\D=\bsigma\cdot(\bp+\bA)$
is nowhere vanishing and that it has a finite, nonzero global
variation lengthscale $L_\bn$  (\ref{eq:Ln}). Let $L\leq L_\bn$,
 then for sufficiently small $\e$ and for any
 $\lambda\ge P=\e^{-4}L^{-2}$ we have
\be
        \frac{\lambda^3}{ (\D^2 + \lambda)^3 }(x,x)
        \leq c \lambda^{1/2}\max\{ B_L^*(x), \lambda\} \, .
\label{eq:cubdiag}
\ee
\end{lemma}
\noindent

For the proof of (\ref{eq:IIres})
 we can use $n(XY^2X; c) = n(YX^2Y; c)$ 
for positive operators $X, Y$, to bring $|V-E|_-$ in the
middle in (\ref{eq:IIdef}),
and estimate it by $V$, then undo it
\be
        (II) \leq \int_0^\infty
          n\Big( V^{1/2} \frac{P^3}{(\D^2+P)^3}V^{1/2} \; ,
        \sfrac{1}{16} E\Big)\rd E = 16P^3 \; \Tr
        \Big( V^{1/2} \, \frac{1}{(\D^2+P)^3} \, V^{1/2}\Big) \; .
\label{eq:IItrace}
\ee
Using  (\ref{eq:cubdiag}) for $\lambda:=P$, we immediately obtain
(\ref{eq:IIres}).
Using that $\Pi_{\tD^2\leq \lambda}\leq 8\lambda^3(\tD^2+\lambda)^{-3}$
from the spectral theorem, we also obtain Corollary \ref{corr:zero}.
Note that (\ref{eq:corr}) holds for any $\lambda\geq0$ since
the case $\lambda<\lambda_*$ is trivial.

\medskip

{\it Proof of Lemma \ref{lemma:cubeloc}.}
  We insert the partition of unity, recall
the bounds (\ref{eq:fin})  and use (\ref{eq:deqd}) to obtain
\bey
        (\D^2 + \lambda)^3 &=&
        \sum_{k\in\Lambda} \Big[ \D^3\theta_k^2 \D^3 +
        3 \lambda\D^2 \theta_k^2 \D^2
        + 3 \lambda^2 \D \theta_k^2 \D   + \lambda^3\theta_k^2\Big]\cr
        &\ge& c \sum_{k\in\Lambda} \Big[ \D^3_k\theta_k^2 \D^3_k +
        \lambda\D_k^2
        \eta_k^2 \D^2_k + \lambda^2\D_k \omega_k^2 \D_k
        +\lambda^2\D_k \theta_k^2 \D_k + \lambda^3\chi_k^2
         \Big]\; .
\label{part}
\eey
We use the
$\lambda^2 \D_k \theta_k^2 \D_k$ term to save
a derivative in the third direction. Let $\wt \D_k$ be the Dirac
operator constructed in Section \ref{sec:bdd}  for the
vector potential $\bA_k$.
Then there exists a spin rotation $U_k:\bR^3\mapsto SU(2)$ (see
(\ref{eq:unit}))
 such that
\be
        \D_k = U_k \wt \D_k U_k^* \; .
\label{eq:unitk}
\ee

 By Schwarz' inequality,
$|\nabla_k\theta|\leq c(\e L)^{-1}$ and  (\ref{eq:third}) from
Lemma~\ref{lemma:lower}
applied to the operator $\wt\D_k^2$ we have
$$
        \D_k \theta_k^2 \D_k \ge c \theta_k \D_k^2 \theta_k
        - c(\e L)^{-2}\eta_k
         \ge c \theta_k U_k D_{\bn_k}^2 U_k^* \theta_k
        - c\e^2 P\eta_k \; .
$$
We can sum up these inequalities, use
 (\ref{eq:fin}) and $\lambda\ge P$,
 to obtain from (\ref{part}) that for sufficiently small $\e$
\bey
        (\D^2 + \lambda)^3 &\ge&
         c\sum_{k} \Big[ \D^3_k\theta_k^2 \D^3_k + \lambda\D_k^2
        \eta_k^2 \D^2_k+ \lambda^2\D_k    \omega_k^2 \D_k
         + \lambda^3 \chi_k^2\Big]\cr
        &&+     c\lambda^2\sum_k
        \theta_k U_k (D_{\bn_k}^2 +\lambda)U_k^*\theta_k \; . \nonumber
\eey
Applying the estimate (\ref{eq:cube})
from Proposition \ref{lemma:lower} to the operators $\wt\D_k = U_k^*\D_k U_k$
 we obtain that
$$
                (\D^2 + \lambda)^3 \ge
        \sum_k
        \theta_k  U_k\Big( (\delta_k T_k)^3 + \lambda^2 D_{\bn_k}^2
         + \lambda^3\Big) U_k^* \theta_k  \; ,
$$
where $\delta_k:= \min\{\lambda b_k^{-1}, 1\}$,
the operators $\bD_k$ and $T_k$
 are obtained by applying the construction of Section \ref{sec:bdd}
for the vector potential $\bA_k$.

Hence by the ``Pull-up'' formula (\ref{eq:pullup}) we get
\be
        \frac{1}{ (\D^2 + \lambda)^3 }
        \leq c\, \sum_{k\in\Lambda} \theta_k U_k \, \frac{1}{ (\delta_kT_k)^3 +
         \lambda^2 D_{\bn_k}^2 + \lambda^3}  \, U_k^*
        \theta_k \; .
\label{eq:cubloc}
\ee
We  apply (\ref{eq:diag}) for the magnetic field $\bB_k=\nabla\times\bA_k$
satisfying the bound (\ref{eq:bb}) with $b=cb_k$ from (\ref{eq:bkbound})
to get
$$
        \frac{1}{ (\D^2 + \lambda)^3 }(x,x)
        \leq c\, \lambda^{1/2} \sum_{k\, : \, \theta_k(x)\neq 0}
        \max\{ b_k, \lambda\} \; ,
$$
 and use that only finitely
many $\theta_k$'s are nonzero at any $x\in\bR^3$, and
for all such $k$'s  we have $b_k\leq B_L^*(x)$. This completes the
proof of (\ref{eq:cubdiag}). $\;\;\Box$

\section{Proof of the positive energy regime}\label{sec:res}
\setcounter{equation}{0}

In this section we prove (\ref{eq:Ires}).
We start by applying the localization (\ref{eq:pullup1}) in
(\ref{eq:Idef})
$$
        (I)\leq  \int_0^\infty
          n\Big( \sum_{k\in \Lambda}
        |V-E|_-^{1/2} \theta_k \frac{1}{\D^2_k+P+E}\theta_k
        |V-E|_-^{1/2} \; , \sfrac{1}{4}\Big)\rd E \; ;
$$
Since the supports of at most $C$ of the functions $\theta_k$ can
 overlap (see (\ref{eq:fin})), we can ``pull out''
the summation
at the expense of descreasing  the energy threshold $\sfrac{1}{4}$
(see Section 6 of \cite{ES-IV} for more details). Therefore
\be
        (I)\leq \sum_{k\in \Lambda}\int_0^\infty
          n\Big( |V-E|_-^{1/2} \theta_k \frac{1}{\D^2_k+P+E}\theta_k
        |V-E|_-^{1/2} \; , \sfrac{1}{4(C+1)}\Big)\rd E \; .
\label{eq:I1}
\ee
For any $\lambda\in\bR$ let
$$
        \Pi_k(\lambda): = \chi(\D^2_k \leq \lambda)
$$
denote
  the spectral projection below the level $\lambda$
in the spectrum of $\D^2_k$. We have
$$
        (I)\leq (I/1) + (I/2)
$$
with
\be
        (I/1) := \sum_{k\in \Lambda}\int_0^\infty
          n\Big( |V-E|_-^{1/2} \theta_k
        \frac{I-\Pi_k( b_k-P-E) }{\D^2_k+P+E}\theta_k
        |V-E|_-^{1/2} \; , \sfrac{1}{8(C+1)}\Big)\rd E  \; ,
\label{eq:I1est}
\ee
\be
        (I/2) := \sum_{k\in \Lambda}\int_0^\infty
          n\Big( |V-E|_-^{1/2} \theta_k
        \frac{\Pi_k( b_k-P-E) }{\D^2_k+P+E}\theta_k
        |V-E|_-^{1/2} \; , \sfrac{1}{8(C+1)}\Big)\rd E  \; .
\label{eq:I2est}
\ee

The estimate of the first term is easy:
$$
        \frac{I-\Pi_k( b_k-P-E) }{\D^2_k+P+E}
        \leq \frac{c}{\D^2_k+cb_k+E} \leq  \frac{c}{(-i\nabla +\bA_k)^2+E}
$$
using the spectral theorem first, then
the Lichnerowicz formula in the Euclidean coordinates:
$$
        \D^2_k = (-i\nabla +\bA_k)^2 + \bsigma\cdot \bB_k
$$
and $\bsigma\cdot \bB_k\ge -cb_k$, where the constant is from
(\ref{eq:bkbound}).
Then we can proceed as in the proof of the original Lieb-Thirring inequality
\cite{LT} for each summand in (\ref{eq:I1est}):
\bey\lefteqn{
      \int_0^\infty  n\Big( |V-E|_-^{1/2} \theta_k
        \frac{c}{(-i\nabla +\bA_k)^2+E}\theta_k
        |V-E|_-^{1/2} \; , \sfrac{1}{8(C+1)}\Big)\rd E}
        \qquad\qquad\qquad\cr
        &\leq& 64(C+1)^2\int_0^\infty\Tr \Big( |V-E|_-^{1/2} \theta_k
        \frac{c}{(-i\nabla +\bA_k)^2+E}\theta_k
        |V-E|_-^{1/2} \Big)^2\rd E\cr
        &\leq& c\int_0^\infty\Tr \Big( |V-E|_-^2 \theta_k^4
        \Big[\frac{1}{(-i\nabla +\bA_k)^2+E}\Big]^2\Big)\rd E\cr
        &\leq&  c\int_0^\infty\int|V(x)-E|_-^2 \theta_k^4(x)
        \Big[\frac{1}{-\Delta+E}\Big]^2(x,x) \rd x\rd E\cr
        &\leq& c\int V^{5/2}\chi_k \; , \nonumber
\eey
where we also
used the diamagnetic inequality for the square of the resolvent.
After the summation over $k$ and using (\ref{eq:fin}), we obtain
the second term in the estimate (\ref{eq:Ires}).

\bigskip

For the term (I/2), we first notice that the $\rd E$ integration
can be restricted to $E\leq b_k$, and we can also assume that
 that $P\leq b_k$. After this restriction
we estimate $\Pi_k( b_k-P-E)$ by $\Pi_k( b_k)$
and we obtain
\bey
        (I/2)&\leq& \sum_{k\in \Lambda}\int_0^{b_k}
          n\Big( |V-E|_-^{1/2} \theta_k
        \frac{\Pi_k( b_k) }{\D^2_k+P+E}\theta_k
        |V-E|_-^{1/2} \; , \sfrac{1}{8(C+1)}\Big)\rd E \cr
        &\leq&  8(C+1)\sum_{k\in \Lambda}\int_0^{b_k}
        \Tr\Big( |V-E|_-^{1/2} \theta_k
        \frac{\Pi_k( b_k) }{\D^2_k+P+E}\theta_k
        |V-E|_-^{1/2}\Big)  \rd E\cr
        &\leq& c \sum_{k\in \Lambda}\int_0^{b_k} \!\!\int_{\bR^3}
         |V(x)-E|_- \theta_k^2(x) \Big[ \int_0^\infty
        \frac{\chi(\lambda\leq b_k) }{\lambda+P+E}\; \rd \Pi_k(\lambda)
        \Big](x,x)\, \rd x \,  \rd E \; ,
\label{eq:I2}
\eey
using the spectral resolution of $\D_k^2$.
Clearly
\bey\lefteqn{
   \int_0^\infty   \frac{\chi(\lambda\leq b_k )}{\lambda+P+E}
    \rd\Pi_k(\lambda) }\qquad\qquad\cr
        &=& \frac{1}{b_k+P+E}\Pi_k(b_k) -\frac{1}{P+E}\Pi_k(0)
    + \int_0^{b_k} \frac{1}{(\lambda+P+E)^2}\Pi_k(\lambda) \rd \lambda\cr
  &\leq& \frac{1}{b_k+E}\Pi_k(b_k)+ \frac{P}{(P+E)^2} \Pi_k(P)
  + \int_P^{b_k} \frac{1}{(\lambda+E)^2}
  \Pi_k(\lambda) \rd \lambda  \; .
\label{eq:intpart}
\eey
The conclusion of Corollary \ref{corr:zero} also applies to $\D_k$,
replacing the variation lengthscale $L_\bn$ with $\e L_\bn$, see (\ref{eq:Lnk}),
therefore
\be
        \Pi_k (\lambda)(x,x)\leq c\lambda^{1/2}\max\{ B_L^*(x), \lambda\} \;
\label{eq:densk}
\ee
for $\lambda\ge \e^{-6}L^{-2}$, but then the same inequality also holds
for all $\lambda\ge P=\e^{-4}L^{-2}$ if $c$ is changed to $c\e^{-3}$
using the monotonicity of $\lambda\mapsto\Pi_k(\lambda)$.

{F}rom (\ref{eq:intpart}), (\ref{eq:densk}) and the fact that $P\leq b_k\leq
B^*_L(x)$ for $x\in\mbox{supp}\, (\theta_k)$,
we obtain
\bey
    \Bigg[\int_0^\infty   \frac{\chi(\lambda\leq b_k )}{\lambda+P+E}
    \rd\Pi_k(\lambda) \Bigg](x,x) &\leq& c\e^{-3} B_L^*(x)\Bigg(
        \frac{b^{1/2}_k}{b_k+E}
    + \frac{P^{3/2}  }{(P+E)^2} + \int_P^{b_k} \!
    \frac{\lambda^{1/2}\rd \lambda}{(\lambda+E)^2}  \Bigg)\cr
    &\leq & c\e^{-3}E^{-1/2}B_L^*(x) \; . \nonumber
\eey
Therefore we can complete the estimate of $(I/2)$ from
(\ref{eq:I2})
$$
   (I/2)\leq c\e^{-3} \sum_{k\in \Lambda}
        \int_0^\infty \int_{\bR^3}B_L^*(x) \theta_k^2(x)
        |V(x)-E|_- E^{-1/2}
       \rd x\; \rd E = c\e^{-3} \int B_L^* V^{3/2} \; .
$$
This completes the proof of (\ref{eq:Ires})
after choosing $\e$ to be a sufficiently small universal constant.$\;\;\Box$

\appendix

\section{Proof of the technical lemmas}
\setcounter{equation}{0}

\subsection{Proof of Lemma \ref{lemma:ext}} \label{sec:extproof}

 Let $\cP_k$ be the plane
through $k\in \Lambda$ and orthogonal to $\bn(k)$.
 We define $\wt\bn_k: = \chi_k \bn + (1-\chi_k)\bn(k)$.
Since $\wt\bn_k - \bn(k) = (\bn -\bn(k))\chi_k=O(\e)$ from (\ref{eq:Ln}),
we have $\wt\bn_k\cdot \bn(k)\ge
1-O(\e)$ and $|\wt\bn_k|=1+O(\e)$,
in particular $\wt\bn_k\neq 0$ for sufficiently small $\e$.
Moreover $\nabla \wt\bn_k = \chi_k \nabla \bn + (\bn - \bn(k))
\nabla\chi_k$, therefore $\nabla\wt\bn_k = O(L^{-1})$
since $|\bn - \bn(k)|=O(\e)$ on the support of $\chi_k$
from (\ref{eq:Ln})
and $L\leq L_\bn$. Similar bounds hold for the higher derivatives.
We  define $\bn_k:= \wt\bn_k/|\wt\bn_k|$ and
(\ref{eq:Lnk}) can be easily checked.

Now $\bB_k$ is defined as $\bB_k:= f_k \bn_k$ where the function
$f_k$ is constructed as follows.
We define
$$
        f_k: = \chi_k |\bB| + (1-\chi_k)|\bB(k)|  \quad \mbox{on} \; \cP_k\; ,
$$
then clearly $|f_k|\leq cb_k$ on $\cP_k$.

Since the vectorfield $\bn_k$ is nearly parallel (clearly
$\bn_k\cdot \bn(k)\ge 1-O(\e)$),
its integral curves define a transversal foliation
to the plane $\cP_k$. Hence we can extend $f_k$ from $\cP_k$ onto the
whole $\bR^3$
by ensuring that $\bB_k$ is divergence free, i.e.,
 $\mbox{div} \; \bB_k = f_k\mbox{div}\; \bn_k + \nabla_{\bn_k}
f_k =0$.  This requires integrating the equation
$$
        \nabla_{\bn_k} (\log f_k) = - \mbox{div}\; \bn_k
$$
 along the
integral curves of $\bn_k$. Since $| \nabla\bn_k|\leq c L^{-1}$,
and it vanishes outside of $\Omega_k^\#$, we obtain $|f_k|
 +L|\nabla (f_k\bn_k)\leq cb_k$ everywhere as required in (\ref{eq:bkbound}).

The original field strength $|\bB|$ also satisfies the equation
$$
        \nabla_{\bn} (\log |\bB|) =  - \mbox{div}\; \bn,
$$
and since
$\bn_k = \bn$ on $\Omega_k$ by definition and $|\bB| = f_k$ on
$\cP_k\cap \Omega_k^*$,
we obtain that $f_k = |\bB|$ on $\Omega_k$. Therefore $\bB_k\equiv \bB$
on $\Omega_k^*$.

This completes the definition of the extension $\bB_k:= f_k \bn_k$
of the field $\bB$ from $\Omega_k^*$ to the whole space together with
the estimates on the field strength (\ref{eq:bkbound})
and on the variation of the field direction (\ref{eq:Lnk}).

Finally, the appropriate vector potential $\bA_k$ is defined
as $\bA_k : = \bA + \bA_k^\#$, where $\bA_k^\#$ is the Poincar\'e
gauge of the magnetic field $\bB_k -\bB$ centered at $k$. Since
$\bB_k \equiv \bB$ on $\Omega_k^*$ and $\Omega_k^*$
 is convex, we have $\bA_k^\#\equiv 0$
on $\Omega_k^*$ as well.
This completes the proof of Lemma \ref{lemma:ext}. $\;\;\Box$

\subsection{Proof of Lemma \ref{lemma:comm}}\label{sec:comm}

Using (\ref{eq:commD}) and $D_\bn= D_3$,
we  compute the commutator:
$$
    [ \bD^t Z \bD , D_3] = [D_j Z_{jk} D_k, D_3]
    = D_j Z_{jk} [D_k, D_3] +i D_j(\partial_3 Z_{jk})  D_k
   +[D_j, D_3] Z_{jk} D_k
$$
$$
    = (-i)\Bigg( D_j Z_{jk}  \Big(
     Q_{k3}^m   D_m  -\sfrac{i}{2}(\partial_m Q_{k3}^m) + \beta_{ k3}\Big)
         + \Big(
     Q_{j3}^m   D_m   -\sfrac{i}{2}(\partial_m Q_{j3}^m) + \beta_{ j3}
     \Big)Z_{jk} D_k   -  D_j(\partial_3 Z_{jk})  D_k\Bigg)
$$
$$
    =  (-i)\Bigg( D_j Z_{jk} Q_{k3}^m   D_m
     - \sfrac{i}{2}(\partial_m Q_{j3}^m) (Z_{kj}-Z_{jk})D_k
     -\sfrac{1}{2} \partial_j(Z_{jk}(\partial_m Q_{k3}^m))
     +  Z_{jk}   \beta_{ k3} D_j
     -i \partial_j( Z_{jk}   \beta_{ k3})
$$
$$
     + \Big(
     D_mQ_{j3}^m  + \beta_{ j3}
     \Big)Z_{jk} D_k
     -  D_j(\partial_3 Z_{jk})  D_k\Bigg)\; ,
$$
and
$$
      [h, D_3]= i\partial_3 h\; .
$$

Now we match the coefficients in
$$
     [ \bD^t Z \bD + h , D_3] =0 \; .
$$
Matching the highest (second) order terms gives
$$
    \partial_3 Z = ZQ +Q^tZ
$$
with the matrix $Q_{ab}:= Q_{a3}^b$.

On the supporting plane we choose $Z\equiv I$.
Using (\ref{eq:ch}) we have
 $\nabla^\gamma Q = O(\e^{-\gamma}L^{-1-\gamma})$, $0\leq \gamma\leq 4$,
 in a neighborhood of size $O(\e L)$,
 otherwise $Q\equiv 0$, hence
we obtain the solution $Z$ satisfying (\ref{eq:ZI}), and $Z$ is
clearly a symmetric, real matrix.

\bigskip

Using the established symmetry of $Z$,
to match the first order terms, we need that
$$
      Z_{kj}\beta_{j3} =0
    \qquad k=1,2,3
$$
but clearly $\beta_{j3} =0$ because $e_3=\bn$ was the
direction of the magnetic field.

Finally the constant term can be matched by choosing $h$ to be 0
on the supporting plane and such that
$h$ satisfy
$$
   \partial_3 h = -\sfrac{1}{2} \partial_j(Z_{jk}(\partial_m Q_{k3}^m))\; .
$$
Using again that $\nabla^\gamma Q = O(\e^{-\gamma}
L^{-1-\gamma})$, $0\leq \gamma\leq 4$,
 in a neighborhood of size $O(\e L)$, otherwise $Q\equiv 0$, we obtain
that the solution satisfies (\ref{eq:hest}). $\;\;\Box$

\subsection{Proof of Lemma \ref{lemma:lower}}\label{sec:lowproof}

{\it Proof of  (\ref{eq:third}).}
We use $\wt\D= \bsigma\cdot \bPi$, (\ref{eq:pid}),
the self-adjointness of $D_j$, (\ref{eq:ch})
and a Schwarz' inequality  to obtain
\be
        \wt\D^2 = \Big( \sum_{j=1}^3 \si^j \big[ D_j +\sfrac{i}{2}\mbox{div}\;
        e_j+\sfrac{1}{2}  \sigma\cdot \bomega(e_j)
        \big] \Big)^2
        \ge \frac{1}{2} \Big( \sum_{j=1}^3 \si^j  D_j\Big)^2 - O(L^{-2}) \; .
\label{eq:1}
\ee

Furthermore
\be
        \Big( \sum_{j=1}^3 \si^j  D_j\Big)^2
        = (\sigma_\perp\cdot D_\perp)^2 + D_3^2
        + \sum_{j=1}^2 \{\si^j  D_j, \si^3 D_3\} \; ,
\label{eq:2}
\ee
where $\sigma_\perp\!\cdot\! D_\perp:= \si^1D_1+\si^2D_2$ and
 $\{X, Y\}= XY+YX$ denotes the anticommutator. Using that
$\si^1\si^3=-i\si^2$ and $\si^2\si^3=i\si^1$, we obtain
from (\ref{eq:commD}) and  $\beta(e_3, e_j)=0$, $j=1,2$,  that
\bey
        \{\si_\perp\!\cdot\! D_\perp, \si^3 D_3\}
        &=& i\si^2[D_3, D_1] -i\si^1[D_3,  D_2] \cr
        &=&\sum_{a,b=1}^2\si^a M_{ab}D_b + D_b M_{ab}
        \si^a + \sfrac{1}{2}\sum_{a=1}^2 (-1)^a \si^a( Q_{3,\bar a}^3 D_3
        + D_3 Q_{3\bar a}^3)
\label{eq:3}
\eey
with $M_{ab} :=\sfrac{1}{2} (-1)^a Q_{3, \bar a }^b $,
$a, b =1,2$, where $\bar 1:=2$ and $ \bar 2:=1$.
We use Schwarz' inequality and that $Q_{jk}^\ell = O(L^{-1})$ to estimate
the last terms containing $D_3$.
In summary, we obtain from (\ref{eq:1})--(\ref{eq:3}) that
\be
        \wt\D^2 \ge  \frac{1}{2}\Big[
        (\sigma_\perp\cdot D_\perp)^2 + \sfrac{1}{2}
        D_3^2 + \si^a M_{ab}D_b + D_b M_{ab}
        \si^a\Big] - O(L^{-2})\; .
\label{eq:sch}
\ee

We  compute  a two dimensional Lichnerowicz formula
$$
                (\sigma_\perp\!\cdot\! D_\perp)^2
        = D_1^2 + D_2^2 +i\si^3[D_1, D_2]
        = D_1^2 + D_2^2 + B\si^3 +  \sfrac{1}{2}\si^3
        (Q_{12}^m D_m + D_m Q_{12}^m)
$$
using that $\beta_{12}= B$.
Let $P_\pm: =\sfrac{1}{2}(1\pm \si^3)$ and note that $P_-$, $P_+$ are
orthogonal projections commuting with $\si^3$, and
 $P_\pm \si^a = \si^a P_\mp$
for $a=1,2$. In particular $(\sigma_\perp\!\cdot\! D_\perp)^2$ commutes
with $P_\pm$.
Therefore
$$
                (\sigma_\perp\!\cdot\! D_\perp)^2 = P_+
                (\sigma_\perp\!\cdot \!D_\perp)^2 P_+
                +P_-
                (\sigma_\perp\!\cdot \!D_\perp)^2 P_-
        \ge P_+ (\sigma_\perp\!\cdot \!D_\perp)^2 P_+
$$
and
\bey
         P_+ (\sigma_\perp\!\cdot \!D_\perp)^2 P_+
        &=& P_+(D_1^2 + D_2^2 + B)P_+ + \sfrac{1}{2} P_+ \si^3
        (Q_{12}^m D_m + D_m Q_{12}^m)P_+\cr
        &&\cr
        &\ge& P_+(D_1^2 + D_2^2 + B)P_+
        - \sfrac{1}{4} P_+(D_1^2 + D_2^2 + D_3^2
        )P_+ - O(L^{-2}) \nonumber
\eey
by a Schwarz' inequality  and (\ref{eq:ch}). We  obtain
from (\ref{eq:sch}) that
\be
        \D^2 \ge \sfrac{1}{8}\Big[ P_+(D_1^2 + D_2^2 + B)P_+ + D_3^2\Big] +
        \sfrac{1}{2} \Big[\si^a M_{ab}D_b + D_b M_{ab}
        \si^a\Big] - O(L^{-2})\; .
\label{eq:sch2}
\ee

Finally, we compute
\bey
     \si^a M_{ab}D_b + D_b M_{ab}\si^a
        &=&  P_+ \Big[ \si^a M_{ab}D_b + D_b M_{ab}\si^a\Big]P_-
        + P_- \Big[ \si^a M_{ab}D_b + D_b M_{ab}\si^a\Big]P_+ \cr
        &=&  2P_- \si^a M_{ab}D_b P_+ + 2P_+ D_b  M_{ab} \si^a P_-\cr 
        &&+ P_+ \si^a [M_{ab}, D_b]P_- + P_- [D_b, M_{ab}]\si^a P_+\; .
\nonumber
\eey
 By a Schwarz' inequality  and (\ref{eq:ch}) we obtain
$$
         \si^a M_{ab}D_b + D_b M_{ab}\si^a \ge - \frac{1}{4}P_+( D_1^2 + D_2^2)
        P_+- O(\e^{-1}L^{-2}) \; ,
$$
which, combined with (\ref{eq:sch2}) gives (\ref{eq:third})
since $D_3=D_\bn$.

\bigskip

\noindent
{\it Proof of (\ref{eq:cube}).}
{F}rom the Lichnerowicz formula (\ref{eq:lich}) and (\ref{eq:pid}) we have
\bey
        \tD^2 &=&
        \Big( D_j -\sfrac{i}{2}\, \mbox{div} \, e_j
         +\sfrac{1}{2}\bsigma\cdot \bomega(e_j)\Big)
        \Big( D_j +\sfrac{i}{2}\, \mbox{div} \, e_j
         +\sfrac{1}{2}\bsigma\cdot \bomega(e_j)\Big)
        + \sigma(\star \beta)\cr
        &=& \bD^2 + \frac{1}{2}[\bsigma\cdot \bomega(e_j) D_j +
        D_j \bsigma\cdot \bomega(e_j)] + \cE \; ,
\label{eq:lichuj}
\eey
where $\cE$ is a zeroth order error that satisfies
\be
        |\cE|+L|\nabla\cE|\leq (P+b)
\label{eq:ce}
\ee
  from (\ref{eq:ch}) and (\ref{eq:bb}).

\begin{lemma} The following comparison estimates hold
\be
        \sfrac{1}{2} \bD^2 - c(P + b)
        \leq \tD^2 \leq 2 \bD^2 + c(P + b) \; ,
\label{eq:dd}
\ee
\be
        \sfrac{1}{4}(\bD^2)^2 - c(P+b)^2\leq \tD^4
         \leq 4(\bD^2)^2 + c(P+b)^2 \; ,
\label{eq:dddd}
\ee
\be
         \sfrac{1}{4} (\bD^2)^3 - c (P+b)^3 \le \tD^6
        \leq 4(\bD^2)^3 + c (P+b)^3 \; .
\label{eq:cubelo}
\ee
\end{lemma}

{\it Proof.} The proof of (\ref{eq:dd}) is straightforward
by a Schwarz' inequality and estimating the geometric
terms  as
\be
        (\bsigma\cdot \bomega(e_j))^2\leq |\bomega(e_j)|^2 \leq \e^2 P
\label{eq:geom}
\ee
from (\ref{eq:ch}). For (\ref{eq:dddd}) we again use a
 Schwarz' inequality
$$
        \tD^4\leq 3(\bD^2)^2 + 3 \Big[D_j (\bsigma\cdot \bomega(e_j))^2D_j+
         \bsigma\cdot \bomega(e_j) D_j^2
        \bsigma\cdot \bomega(e_j) \Big]
         + 3(P+b)^2
$$
The first term in the square bracket is bounded by $\e^2 P \bD^2$
using (\ref{eq:geom}).
In the second term we commute through and obtain
$$
        \bsigma\cdot \bomega(e_j) D_j^2
        \bsigma\cdot \bomega(e_j) \leq 2 D_j (\bsigma\cdot \bomega(e_j))^2D_j
        + 2 |\partial_j \bsigma\cdot \bomega(e_j)|^2
        \leq 2\e^2 P\bD^2 + \e^4 P^2
$$
from (\ref{eq:ch}). Altogether, with  a further Schwarz' inequality
we obtain the upper bound in (\ref{eq:dddd}).
A Similar argument gives a lower bound as well.

Finally, to bound the sixth power from below we use (\ref{eq:dd})
\be
        \tD^6 \ge \tD^2( \sfrac{1}{2}\bD^2 - c(P+b))\tD^2 \; .
\label{eq:6}
\ee
For the first term we use (\ref{eq:lichuj}) and Schwarz
$$
        \tD^2\bD^2\tD^2 \ge \frac{1}{2} (\bD^2)^3
        - c \bsigma\cdot\bomega(e_j)D_j \bD^2 D_j\bsigma\cdot\bomega(e_j)
        - cD_j\bsigma\cdot\bomega(e_j) \bD^2\bsigma\cdot\bomega(e_j)D_j
        - c\cE \bD^2\cE
$$
All the terms  $\bsigma\cdot\bomega(e_j)$ can be commuted in the middle
and estimated by $\e P$ from (\ref{eq:ch}). 
Similarly, $D_j$'s can be commuted through $\bD^2$ to estimate 
$$
        \e P D_j\bD^2 D_j\leq 2\e P (\bD^2)^2 + \e P^2 \bD^2 + \e P(P+b)^2
$$
using (\ref{eq:commD}). 
The resulting terms of the form
$\e P(\bD^2)^2$ and $\e P^2 \bD^2$ can be controlled by $2\e (\bD^2)^3 + c\e P^3$.
The last term in (\ref{eq:6})
can be dealt with in the same way after estimating $\tD^4$
by $(\bD^2)^2$ from (\ref{eq:dddd}). For the  term $\cE \bD^2\cE$ we use
$$
        \cE \bD^2 \cE \leq 2D_j \cE^2 D_j +2|\nabla \cE|^2
$$
and the bound (\ref{eq:ce}).

Putting all these estimates together and applying several
Schwarz' inequality, we easily arrive at the lower bound in
(\ref{eq:cubelo}). The upper bound can be proven similarly,
although we will not need it. $\;\;\Box$

\medskip

After these preparations, now turn to the actual proof
of (\ref{eq:cube}).
\be
        \tD^3 \theta^2 \tD^3 \ge \frac{1}{2} \tD^2\theta\tD^2\theta\tD^2
        - 2 \tD^2 |\nabla\theta|^2\tD^2 \ge
        \frac{1}{2} \tD^2\theta\tD^2\theta\tD^2 -
        \e \lambda  \tD^2 \eta^2 \tD^2 \; .
\label{eq:main1}
\ee
The error can be absorbed into the second term on the left of
(\ref{eq:cube}).
We continue with the main term
\be
        \tD^2\theta\tD^2\theta\tD^2\ge \frac{\delta}{4}\tD \theta\tD^4
        \theta\tD -
        \delta\tD\sigma(\rd\theta)\tD^2
        \sigma(\rd\theta)\tD .
\label{eq:main2}
\ee
In the error term we use (\ref{eq:dd}) and $|\nabla\theta|^2\leq \e P\omega^2$
\be
        \delta\tD\sigma(\rd\theta)\tD^2\sigma(\rd\theta)\tD
        \leq c\delta \tD\sigma(\rd\theta)\bD^2\sigma(\rd\theta)\tD
        + \e\delta P (P+b)\tD \omega^2\tD \; .
\label{err}
\ee
The second term can be absorbed into the third term on the  left side
of (\ref{eq:cube}) since $\delta(P+b)\leq 2\lambda$ and $P\leq \lambda$.
In the first term in (\ref{err}) we can eliminate  the $\sigma$'s
using a Schwarz' inequality, $(\partial_j\theta_j)\si^j \bD^2 \si^k(\partial_k\theta)
\leq 3(\partial_j\theta_j)\bD^2(\partial_j\theta_j)$, and the fact that
$\bD^2$ is identity in the spin space, hence it
commutes with the Pauli matrices. After that we can again
use (\ref{eq:dd}) and commute  $\nabla\theta$ inside
\bey
        \delta \tD\sigma(\rd\theta)\bD^2\sigma(\rd\theta)\tD
        &\leq& c \delta \tD (\partial_j\theta)\bD^2(\partial_j\theta)\tD \cr
        &\leq& c\delta \tD (\partial_j\theta)(\tD^2  + P+b) (\partial_j\theta)\tD\cr
        &\leq& c\delta \tD^2 |\nabla\theta|^2\tD^2 +
        c\delta \tD |\nabla^2\theta|^2\tD +
        c\delta \e P(P+b)\tD\omega^2\tD\cr
        &\leq&  c\e \lambda \tD^2\eta^2 \tD^2 + c\e \lambda^2 \tD \omega^2\tD\; ,
\label{err2}
\eey
and these terms can be absorbed into the lower order
terms on the  left side of (\ref{eq:cube}).

Finally, we continue the estimate of the main term in (\ref{eq:main2}):
\be
        \frac{\delta}{4}\tD \theta\tD^4 \theta\tD
        \ge \frac{\delta^2}{4}\tD \theta\tD^4 \theta\tD
        \ge \frac{\delta^2}{8}\theta\tD^6 \theta
        - \delta^2 \sigma(\rd\theta)\tD^4 \sigma(\rd\theta) \; .
\label{eq:main3}
\ee
In the first term we use (\ref{eq:cubelo}) after the estimate $\delta^2\ge
\delta^3$.
In the error term we change $\tD^4$ to $(\bD^2)^2$
using the upper bound in (\ref{eq:dddd}), then we can eliminate the $\sigma$'s,
use the lower bound in (\ref{eq:dddd}) to change $(\bD^2)^2$ back to $\tD^4$
and finally commute $\partial\theta$'s in the middle similarly
to (\ref{err2}). All the error terms can be
absorbed into the lower order
terms on the  left side of (\ref{eq:cube}). We omit the details.

Putting all these estimates together, we obtain for sufficiently small $\e$
$$
        \wt\D^3 \theta^2 \wt\D^3 + \lambda\wt\D^2\eta^2 \wt\D^2
        + \lambda^2\wt\D\omega^2 \wt\D+\lambda^3\chi^2
          \ge c \, \delta^3 \theta (\bD^2)^3 \theta \; ,
$$
Since $T = \bD\cdot Z\bD + h$ and $Z$, $h$ satisfy (\ref{eq:ZI}),
(\ref{eq:hest}), in particular $Z$ is close to the identity,
 it is a trivial exercise with commutators
to verify that
$$
        (\bD^2)^3 \ge c T^3 - \e P^3
$$
and the error can be absorbed into $\lambda^3\chi^2$.
This completes the proof of (\ref{eq:cube}).
$\;\;\Box$

\subsection{Proof of Lemma \ref{lemma:Tdiag}}\label{sec:Tdiagproof}

Since
$X:=\delta\lambda^{-1}  T$ and $Y:= \lambda^{-1} D_\bn^2$
commute, we can estimate
$$
         \frac{\lambda^3}{ (\delta T )^3 + \lambda^2 D_{\bn}^2 + \lambda^3}
        \leq \frac{c}{ (X + Y^{1/3}  + I)^3}
        \leq c \int_0^\infty t^2 e^{-t} e^{-tX}e^{-tY^{1/3}} \rd t\; ,
$$
therefore
\be
        \frac{\lambda^3}{ (\delta T )^3 + \lambda^2 D_{\bn}^2 + \lambda^3}(x,x)
        \leq c \int_0^\infty t^2 e^{-t}\,  \Big| e^{-tX}(x,y)\Big|
        \; \Big| e^{-tY^{1/3}}(y,x)\Big|  \rd y\;\rd t\; .
\label{eq:4}
\ee
We recall that $\wh T$ and $\wh D_j$
refer to the nonmagnetic counterparts of $T$ and $D_j$, we also
denote $\wh X := \delta\lambda^{-1}  \wh T$, $\wh Y: = \lambda^{-1} \wh
D_\bn^2$. The operator
$T$ is a second order uniformly elliptic operator, its heat kernel
has a Feynman-Kac representation with an
imaginary term in the exponent due to $\alpha$. Following the
standard proof of the diamagnetic inequality using Feynman-Kac formula
(see e.g. \cite{S79}), we can estimate
the oscillatory term trivially by one and  we obtain
$$
        \Big| e^{-tX}(x,y)\Big| \leq e^{-t\wh X}(x,y) \; .
$$
{F}rom standard off-diagonal
elliptic heat kernel estimates (see, e.g. \cite{LiY}) we obtain
\be
         e^{-t\wh X}(x,y) \leq c (t\delta\lambda^{-1})^{-3/2}
        \exp\Big( -\frac{ c (x-y)^2}{t\delta\lambda^{-1}}\Big) \; .
\label{eq:heat}
\ee

The operator $D_\bn$ and $\wh D_\bn$ are conjugated by a phase
factor
$$
        D_\bn = e^{-i\phi}\wh D_\bn e^{i\phi}\; ,
$$
where $\phi$ is a solution of $\partial_\bn \phi  = \a(\bn)$.
Since the integral curves of $\bn$ are disjoint,  a solution
$\phi$ exists globally. Therefore
$$
        Y = e^{-i\phi} \wh Y e^{i\phi}
$$
hence, by spectral theorem,
$$
        \Big| e^{-tY^{1/3}}(y,x)\Big| = \Big| e^{-t\wh Y^{1/3}}(y,x)\Big| \; .
$$
Since $\wh Y$ is a one-dimensional Laplacian, its spectral
decomposition is exactly computable in appropriate coordinates
 $(\xi_1, \xi_2, \xi_3)$ chosen  as follows.
Let $\xi_3$ be the arclength parameter along the integral curves
of $\bn$ with $\xi_3 \equiv 0$ on the supporting plane $\cP$.
We choose an arbitrary Euclidean coordinatization $\xi_\perp:=
(\xi_1, \xi_2)$ on $\cP$.
Since the integral curves give a regular foliation of $\bR^3$,
we can extend the coordinates $\xi_1, \xi_2$ onto $\bR^3$
by setting it constant along each integral curve.

Since the integral curves are straight lines outside of $\Omega^\#$,
and they are almost parallel inside, it is easy to see that
\be
        c_1\leq\frac{ |\xi(x)-\xi(y)|}{|x-y|}\leq c_2
\label{eq:comp}
\ee
for any $x,y$ with universal constants $c_1, c_2$.
A detailed proof of this obvious
fact can be obtained along the lines of the proof of
 Lemma 8.4 of \cite{ES-IV}, therefore we omit it here.
Moreover, the Jacobian determinant $\det({D\xi(x)})$
of the transformation $x\mapsto \xi(x)$ is universally bounded and
likewise for the inverse map.

Define the unitary transform $U:L^2(\bR^3,\rd\xi)\to L^2(\bR^3,\rd x)$
$$
        \left[U\psi\right](x)=|\det({D\xi(x)})|^{1/2}\psi(\xi(x)).
$$
We then have that
$$
         U^*\widehat D_\bn U=-i\partial_{\xi_3}.
$$
Hence
\beys
        e^{-t\widehat{Y}^{1/3}}(x,y)&=&|\det({D\xi(x)})|^{1/2}
        |\det({D\xi(y)})|^{1/2}e^{-t[(-i\partial_{\xi_3})^2]^{1/3}}(\xi(x),\xi(y))\\
        &=&|\det({D\xi(x)})|^{1/2}
        |\det({D\xi(y)})|^{1/2}\delta(\xi_\perp(x), \xi_\perp(y))
        \int_\bR e^{-t\lambda^{-1/3}|p|^{2/3}}
        e^{ip(\xi_3(x)-\xi_3(y))} \rd p.
\eeys
Hence
\be
        \Big|   e^{-t\wh Y^{1/3}}(x,y)\Big|
        \leq c \lambda^{1/2}t^{-3/2}\delta(\xi_\perp(x), \xi_\perp(y)) \; .
\label{eq:1dim}
\ee

Now we can estimate the diagonal kernel
(\ref{eq:4}), using
(\ref{eq:heat}), (\ref{eq:1dim}), (\ref{eq:comp}), and the boundedness
of the Jacobian determinants;
\bey\lefteqn{
        \frac{\lambda^3}{ (\delta T )^3 + \lambda^2 D_{\bn}^2 + \lambda^3}
        (x, x)}\qquad\qquad\qquad\qquad \cr
        &\leq& c\lambda^2\delta^{-3/2} \int_0^\infty \!\!\int_{\bR^3}
        t^{-1} e^{-t}\,
        \exp\Big( -\frac{ c (\xi(x)-\eta)^2}{t\delta\lambda^{-1}}\Big)
        \delta(\eta_\perp, \xi_\perp(x))\rd \eta\, \rd  t \cr
        && \cr
        &=& c\lambda^2\delta^{-3/2} \int_0^\infty \!\!\int_\bR
        t^{-1} e^{-t}
        \exp\Big( -\frac{ c (\xi_3(x)-\eta_3)^2}{t\delta\lambda^{-1}}\Big)
        \rd\eta_3\, \rd t \cr
        &=& c\lambda^{3/2} \delta^{-1} \; .
\eey
This completes the proof of (\ref{eq:diag}). $\;\;\Box$

\bigskip
\bigskip
\noindent Addresses of the authors:

\medskip

\noindent  L\'aszl\'o Erd\H os \\
School of Mathematics\\
 GeorgiaTech Atlanta, GA 30332, U.S.A.\\
 lerdos@math.gatech.edu

\medskip
\noindent
After Jul 1, 2003: \\
Mathematisches Institut der LMU \\
Theresienstrasse 39\\
D-80333, M\"unchen, Germany\\
lerdos@mathematik.uni-muenchen.de

\bigskip

\noindent Jan Philip Solovej\\
Department of Mathematics, University of Copenhagen
\\ Universitetsparken 5, DK-2100, Copenhagen, Denmark
\\
 solovej@math.ku.dk

\end{document}